\documentclass[twocolumn]{aastex61}
\usepackage{mydefs}
\usepackage{amsmath}
\usepackage[caption=false]{subfig}

\shorttitle{nbodykit: a massively parallel large-scale structure toolkit}
\shortauthors{Hand \& Feng et al.}

\begin{document}
\newcommand{\YF}[1]{\textbf{YF: #1}}
\title{nbodykit: an open-source, massively parallel toolkit for
        large-scale structure}

\correspondingauthor{Nick Hand}
\email{nhand@berkeley.edu}

\author[0000-0002-8809-3939]{Nick Hand}
\affiliation{Astronomy Department, University of California,
Berkeley, CA 94720, USA}
\affiliation{Berkeley Center for Cosmological Physics,
University of California, Berkeley CA 94720}

\author[0000-0001-5590-0581]{Yu Feng}
\affiliation{Berkeley Center for Cosmological Physics,
University of California, Berkeley CA 94720}

\author[0000-0003-0467-5438]{Florian Beutler}
\affiliation{Institute of Cosmology \& Gravitation, Dennis Sciama Building,
University of Portsmouth, Portsmouth, PO1 3FX, UK}
\affiliation{Lawrence Berkeley National Laboratory,
1 Cyclotron Road, Berkeley, CA 94720, USA}

\author{Yin Li}
\affiliation{Berkeley Center for Cosmological Physics,
University of California, Berkeley CA 94720}
\affiliation{Physics Department, University of California,
Berkeley, CA 94720, USA}
\affiliation{Lawrence Berkeley National Laboratory,
1 Cyclotron Road, Berkeley, CA 94720, USA}
\affiliation{Kavli Institute for the Physics and Mathematics of the Universe
(WPI), UTIAS, The University of Tokyo, Chiba 277--8583, Japan}

\author[0000-0002-1670-2248]{Chirag Modi}
\affiliation{Physics Department, University of California,
Berkeley, CA 94720, USA}
\affiliation{Berkeley Center for Cosmological Physics,
University of California, Berkeley CA 94720}

\author{Uro{\v s} Seljak}
\affiliation{Physics Department, University of California,
Berkeley, CA 94720, USA}
\affiliation{Berkeley Center for Cosmological Physics,
University of California, Berkeley CA 94720}

\author{Zachary Slepian}
\altaffiliation{Einstein Fellow}
\affiliation{Berkeley Center for Cosmological Physics,
University of California, Berkeley CA 94720}
\affiliation{Lawrence Berkeley National Laboratory,
1 Cyclotron Road, Berkeley, CA 94720, USA}

\begin{abstract}

We present \nbkit{}, an open-source, massively parallel Python toolkit
for analyzing large-scale structure (LSS) data. Using Python bindings
of the Message Passing Interface (MPI), we provide parallel implementations
of many commonly used algorithms in LSS. \nbkit{} is
both an interactive and scalable piece of scientific software,
performing well in a supercomputing environment while still taking advantage
of the interactive tools provided by the Python ecosystem.
Existing functionality includes estimators of the power spectrum,
2 and 3-point correlation functions, a Friends-of-Friends grouping
algorithm, mock catalog creation via the halo occupation distribution
technique, and approximate $N$-body simulations via the FastPM scheme. The
package also provides a set of distributed data containers, insulated from
the algorithms themselves, that enable \nbkit{} to provide a unified
treatment of both simulation and observational data sets. \nbkit{} can be
easily deployed in a high performance computing environment, overcoming some of
the traditional difficulties of using Python on supercomputers.
We provide performance benchmarks illustrating the scalability of the software.
The modular, component-based approach of \nbkit{} allows researchers to
easily build complex applications using its tools. The
package is extensively documented at \url{http://nbodykit.readthedocs.io},
which also includes an interactive set of example recipes for new users to
explore. As open-source software, we hope \nbkit{}
provides a common framework for the community to use and develop
in confronting the analysis challenges of future LSS surveys.

\end{abstract}


\section{Introduction}
\label{sec:intro}

The analysis of LSS data sets has played a pivotal role in
establishing the current concordance
paradigm in modern cosmology, the $\lcdm$ model. From the earliest galaxy
surveys \citep{Davis:1985, Maddox:1990}, comparisons between the theoretical
predictions for the distribution of matter in the Universe and observations
have proven to be a valuable tool.
Indeed, LSS observations, in combination with cosmic microwave
background (CMB) measurements, provided some of the earliest evidence for
the $\lcdm$ model, e.g., \citet{Efstathiou:1990, Krauss:1995, Ostriker:1995}.
Interest in LSS surveys increased immensely following the first direct
evidence for cosmic acceleration \citep{Riess:1998,Perlmutter:1999}, as it was
realized that the baryon acoustic oscillation (BAO) feature imprinted on
large-scale clustering provided a ``standard ruler'' to map the expansion history
\citep{Eisenstein:1998, Blake:2003, Seo:2003}. From its first measurements
\citep{Cole:2005, Eisenstein:2005} to more
recent studies \citep{Font-Ribera:2014, Delubac:2015, Alam:2017, Slepian:2017},
the BAO has proved to be a valuable probe of cosmic acceleration, enabling
the most precise measurements of the expansion history of the Universe
over a wide redshift range.
Analyses of these data sets have also pushed us closer to
answering other important questions in contemporary cosmology, including
deviations from General Relativity \citep{Mueller:2016},
the neutrino mass scale \citep{Lesgourgues:2006, Beutler:2014}, and
the existence of primordial non-Gaussianity
\citep{Slosar:2008, Desjacques:2010}.

The foundations of the numerical methods used in LSS data analysis today
go back several decades.
\citet{Hockney:1981} discussed several important computer simulation
methods, including but not limited to mass assignment interpolation windows
and the interlacing technique for reducing aliasing.
The Friends-of-Friends (FOF) algorithm for identifying halos from a
numerical simulation was first utilized in \citet{Davis:1985}.
The most commonly used clustering estimators for the
two-point correlation function (2PCF) and power spectrum were first
developed in \citet{Landy:1993} and \citet{Feldman:1994}, respectively,
and techniques to measure anisotropic clustering via a multipole basis
were first used around the same time, e.g., \citet{Cole:1995}.
Other modern, well-established numerical techniques include
$N$-body simulation methods, e.g., \citet{Springel:2001, Springel:2005},
and the use of KD-trees in correlation function estimators \citep{Moore:2001}.

Recent years have brought important updates to these analysis techniques.
Advances in LSS observations, with increased sample sizes and statistical
precision, have driven the development of new statistical estimators,
while also increasing modeling complexities and creating a need to reduce
wall-clock times. Recently, we have seen faster power spectrum and
2PCF multipole estimators
\citep{Yamamoto:2006,Scoccimarro:2015,Bianchi:2015,Slepian:2015b,Slepian:2016,Hand:2017}
and improved FOF algorithms \citep{Springel:2005, Behroozi:2013,Feng:2017}.
Highly optimized software, e.g., \Corrfunc{} \citep{Sinha:2017},
is also becoming increasingly common.
New statistical estimators, e.g., \citet{Slepian:2015,Slepian:2017b,Castorina:2017},
are being developed to extract as much information as possible
from LSS surveys. The rise of particle mesh simulation methods
\citep{Merz:2005,Tassev:2013,White:2014,Feng:2016b} has offered a
computationally cheaper alternative to running full $N$-body simulations.
Finally, tools have emerged to help deal with the growing
complexities of modeling the connection between halos and galaxies
\citep{Hearin:2017}. These examples represent just a sampling of the
recent updates to LSS data analysis and modeling techniques.

The well-established foundation of LSS numerical methods suggests the
community could benefit from a standard software package providing
implementations of these methods. Such a package would also serve as a
common framework for users as they incorporate future extensions and
advancements. Given the already rising wall-clock times of current analyses
and the expected volume of data from next-generation LSS surveys,
scaling performance should also be a key priority.

Several computing trends in the past few years have emerged to help
make such a software package possible. First, the Python programming
language\footnote{\url{http://python.org}} has emerged as the most popular
language in the field of astronomy \citep{Momcheva:2015, NSF:2017}, and the
\astropy{}\footnote{\url{http://www.astropy.org}} package
\citep{Astropy:2013} has led the development of an
astronomy-focused Python ecosystem. Python's elegant syntax and dynamic
nature make the language easy to learn and work with.
Combined with its object-oriented focus and the larger ecosystem
containing \scipy\footnote{\url{https://www.scipy.org}} \citep{Jones:2001},
\numpy{}\footnote{\url{http://www.numpy.org}} \citep{NumPy:2011},
\ipython{}\footnote{\url{https://ipython.org}} \citep{Perez:2007},
and \jupyter{}\footnote{\url{http://jupyter.org}} \citep{Thomas:2016},
Python is well-suited for both
rapid application development and use in scientific research.
Second, the availability and performance of large-scale computing resources
continues to grow, and initiatives, e.g.,
The Exascale Computing Project,\footnote{\url{https://www.exascaleproject.org}}
have been established to ensure the sustainability of this trend.
At the same time, solutions to the
traditional barriers to using Python on massively parallel,
high-performance computing (HPC) machines have been developed. The
\mpipy{} package \citep{Dalcin:2008, Dalcin:2011} has facilitated the
development of parallel Python applications by providing bindings of
the Message Passing Interface (MPI) standard. Furthermore, tools have
been developed, e.g., \citet{Feng:2016}, to alleviate the start-up
bottleneck encountered when launching Python applications on HPC systems.

Motivated by these recent developments, we present the first
public release of \nbkit{} (v0.3.0), an open-source, parallel toolkit
written in Python for use in the analysis of LSS data. Designed for use
on HPC machines, \nbkit{} includes fully parallel implementations
of a canonical set of LSS algorithms. It also includes a set of
distributed and extensible data containers, which can support
a wide variety of data formats and large volumes of data.
These data containers are insulated from the algorithms themselves, allowing
\nbkit{} to be used for either simulation or observational data sets.
We have balanced the scalable nature of \nbkit{} with an object-oriented,
component-based design that also facilitates interactive use.
This allows researchers to take advantage of interactive Python tools,
e.g., the \jupyter{} notebook, as well as integrate \nbkit{} components
with their own software to build larger applications that solve specific
problems in LSS.

\nbkit{} has been developed, tested, and deployed on the
Edison and Cori Cray supercomputers at the National Energy Research Scientific
Computing Center (NERSC) and has been
utilized in several published research studies
\citep{Hand:2017,Hand:2017b,Ding:2017,Pinol:2017,Schmittful:2017,Modi:2016,Feng:2016b,Waters:2016}.
Since its start, it has been developed on GitHub as open-source
software at \url{https://github.com/bccp/nbodykit}.

The objective of this paper is to
provide an overview of the \nbkit{} software and familiarize the community
with some of its capabilities.
We hope that researchers find \nbkit{}
to be a useful tool in their scientific work and in the spirit of open science,
that it continues to grow via community contributions.
Extensive documentation and tutorials are available at
\url{http://nbodykit.readthedocs.io}, and we do not aim to provide such
detailed documentation in this work. The documentation also
includes instructions for launching an interactive environment containing
a set of example recipes. This allows new users to explore \nbkit{} without
setting up their own \nbkit{} installation.

The paper is organized as follows. We provide a broad overview
of \nbkit{} in Section~\ref{sec:overview} and discuss a more detailed
list of its capabilities in Section~\ref{sec:capabilities}.
We describe our development process and deployment strategy for
\nbkit{} in Section~\ref{sec:dev-approach}.
Section~\ref{sec:inaction} presents an illustrative example use case,
and Section~\ref{sec:benchmarks} outlines performance benchmarks for various
algorithms. Finally, we conclude and summarize in
Section~\ref{sec:conclusions}.

\section{Overview}
\label{sec:overview}

\subsection{Initializing \nbkit{}}

A core design goal of \nbkit{} is maintaining an interactive user experience,
allowing the user to quickly experiment and to prototype new analysis pipelines
while still leveraging the power of parallel processing when necessary.
We adopt a ``lab'' framework for \nbkit{}, where all of the necessary data
containers and algorithms can be imported from the \code{nbodykit.lab}
module. Furthermore, we utilize Python's \code{logging} module to
print messages at runtime, which allows users to track the progress
of the application in real time.
Typically, applications using \nbkit{} begin with the following
statements:

\begin{figure}[h]
\centering
\includegraphics[width=0.7\columnwidth]{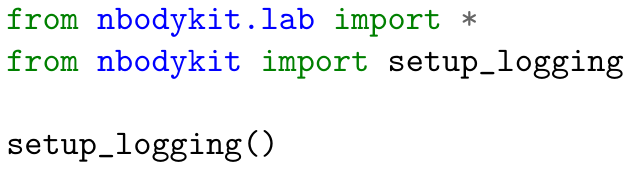}
\end{figure}

\subsection{The \nbkit{} Ecosystem}

\nbkit{} is explicitly maintained as a pure Python package. However, it depends
on several compiled extension packages that each focus on more specialized
tasks. This approach enables \nbkit{} to describe higher-level abstractions
in Python and retains the readability, syntax, and user interface benefits
of the Python language. For computationally expensive sections of the code base,
we use the compiled extension packages for speed.
With the emergence of Python package managers such as
Anaconda,\footnote{\url{https://anaconda.com}} the availability of binary versions
of these compiled packages for different operating systems has sufficiently
eased most installation issues in our experience
(see Section~\ref{sec:installing}).

Below, we describe some of the more important dependencies of
\nbkit{}, each of which is focused on solving a particular problem:

\begin{itemize}
  \item \pfftpython{}: a Python binding of the \code{PFFT} software
  \citep{Pipping:2013}, which computes parallel fast Fourier transforms (FFTs)
  \citep{pfft-python}.
  \item \pmesh{}: particle mesh calculations, including density field
  interpolation and discrete parallel FFTs via \pfftpython{} \citep{pmesh}.
  \item \bigfile{}: a reproducible, massively parallel input/output (IO) library
  for large, hierarchical data sets \citep{bigfile}.
  \item \kdcount{}: spatial indexing operations via KD-trees \citep{kdcount}.
  \item \classylss{}: a Python binding of the \code{CLASS} Boltzmann solver
  \citep{classylss}.
  \item \fastpm{}: a Python implementation of the FastPM scheme for quasi $N$-body
  simulations \citep{fastpm,Feng:2016b}.
  \item \Corrfunc{}: a set of high-performance routines for computing pair counting
  statistics \citep{Sinha:2017}.
  \item \halotools{}: a package to build and test models of the galaxy-halo
  connection \citep{Hearin:2017}.
\end{itemize}

\subsection{A Component-Based Approach}\label{sec:component-approach}

\begin{figure*}
  \includegraphics[width=\textwidth]{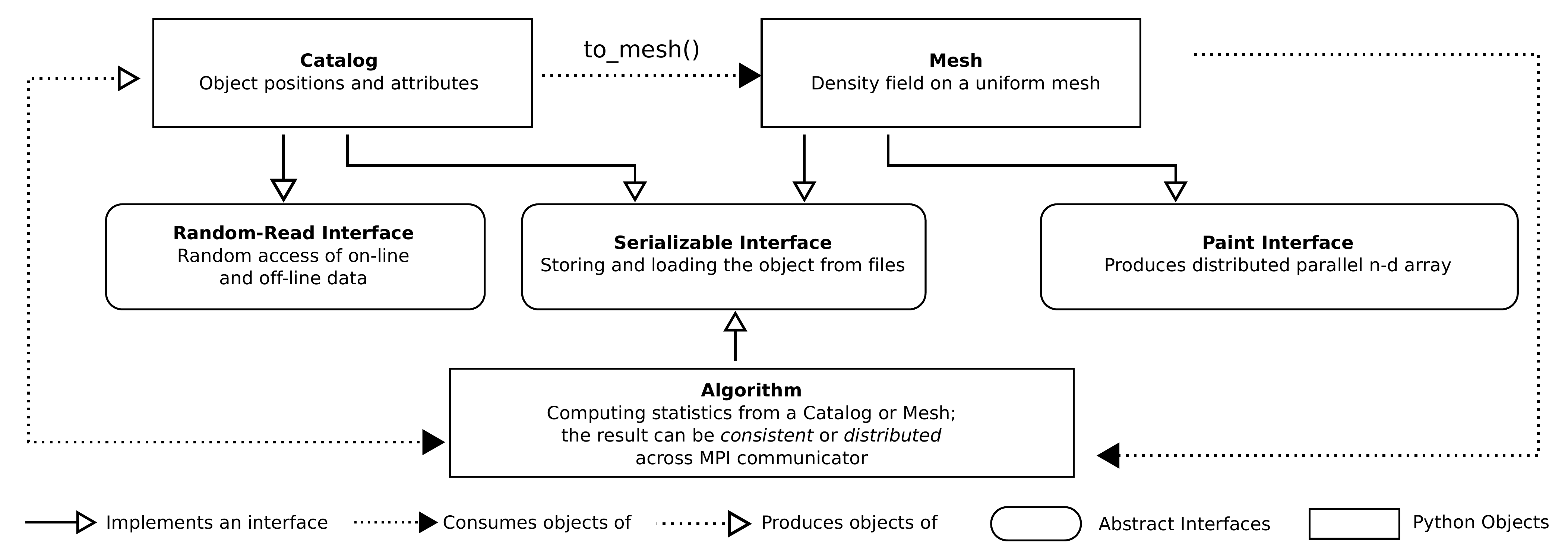}
  \caption
  {
  The components and interfaces of \nbkit{}. The main Python classes
  are Catalog, Mesh, and Algorithm objects, which are described in
  more detail in \S\ref{sec:component-approach}. Algorithm results can
  be \textit{consistent}, where all processes hold the same data, or
  \textit{distributed}, where data is spread out evenly across parallel
  processes.
  }
  \label{fig:interfaces}
\end{figure*}

The design of \nbkit{} focuses on a modular, component-based approach.
The \textit{components} are exposed to the user as a set of Python
classes and functions, and users can combine these components to build their specific applications.
This design differs from the more commonly used alternative in
cosmology software, which is a monolithic application controlled by a
single configuration file, e.g.,
as in \code{CAMB} \citep{Lewis:2000}, \code{CLASS} \citep{Blas:2011},
and \code{Gadget} \citep{Springel:2001}.
We note that modular, object-oriented designs using Python are becoming more
popular recently, e.g., \astropy{}, \yt{} \citep{Turk:2011},
\halotools{} \citep{Hearin:2017}, and Colossus \citep{Diemer:2017}.
During the development process, we have found that a component-based
approach offers greater freedom and flexibility to build complex
applications with \nbkit{}.

We present some of the main classes and interfaces and how data flows through
them in Figure~\ref{fig:interfaces}. In the subsections to follow, we provide an
overview of some of the components outlined in this figure.

\subsubsection{Catalog}

A Catalog is a Python object derived from a \code{CatalogSource}
class that holds information about discrete
objects\footnote{Here, ``object'' can represent galaxies, simulation particles,
mass elements, etc.} in a columnar format. Catalogs implement a
\textit{random-read interface},
which allows users to access arbitrary slices of the data while also taking
advantage of the high throughput of a parallel file system.
Often, users will initialize Catalog objects by reading data from a
file on disk, using a \numpy{} array already stored in
memory, or by generating simulated particles at runtime using one of
\nbkit{}'s built-in classes.

\subsubsection{Mesh}

A Mesh is a Python object that computes a discrete representation
of a continuous quantity on a uniform mesh.  It is
derived from a \code{MeshSource} class and provides a
\textit{paintable} interface, which refers to the process of ``painting''
the density field values onto the discrete mesh cells. When the
user calls the \code{paint()} function, the
mesh data is returned as a three-dimensional array.
Mesh objects can be created directly from a Catalog via the \code{to\_mesh()}
function or by generating simulated fields directly on the mesh.

\subsubsection{Algorithms}

Algorithms are implemented as Python classes and interact with data
by consuming Catalog and Mesh objects as input. The algorithm is executed
when the user initializes the class, and the returned instance stores
the results as attributes.

\subsubsection{Serialization and Reproducibility}

Most objects in \nbkit{} are serializable\footnote{Serialization (and its reverse,
de-serialization) refers to the process of storing a Python object on disk
in a format such that it can be reconstructed at a later time.} via
a \code{save()} function.
Algorithm classes not only save the result of the
algorithm but also save input parameters and meta-data.
They typically implement both a \code{save()} and \code{load()} function, such
that the algorithm result can be de-serialized into an object of the same type.
The two main data containers, catalogs and meshes, can be serialized using
\nbkit{}'s intrinsic format which relies on the massively parallel IO
library \bigfile{} \citep{bigfile}. \nbkit{} includes support for reading
these serialized results from disk back into Catalog or Mesh objects.

\subsection{Parallelism}

\subsubsection{Data-based}

\nbkit{} is fully parallelized using the Python bindings of the MPI standard
available through \mpipy{}. The MPI standard allows processes running in
parallel, each with their own memory, to exchange messages. This mechanism
enables independent results to be computed
by individual processes and then combined into a single result.

Both the Catalog and Mesh objects are \textit{distributed} data containers,
meaning that the data is spread out evenly across the available processes
within an MPI communicator.\footnote{The MPI communicator is responsible for
managing the communication between a set of parallel processes.}
Nearly all algorithm calculations are performed on this distributed data,
with final results computed via a reduce operation across all
processes in the communicator. Rarely throughout the
code base, data is instead gathered to a
single root process, and operations are performed on this data before
re-distributing the results to all processes. This only occurs when
wall-clock time will not be a concern for most use cases and the additional
complexity of a massively parallel implementation is not merited.

The distributed nature of the Catalog object is implemented by using the
random-read interface to access different slices of the tabular data for
different processes. The values of a Mesh object are
stored internally on a three-dimensional \numpy{} array, which is distributed
evenly across all processes. The domain of the 3D mesh is decomposed across
parallel processes using the particle mesh library \pmesh{}, which
also provides an interface for computing parallel FFTs of the mesh data
using \pfftpython{}. The \pfftpython{} software exhibits excellent scaling
with the number of available processes, enabling high-resolution (large number
of cells) mesh calculations.

\subsubsection{Task-based}

The analysis of LSS data often involves hundreds to thousands of
repetitions of a single, less computationally expensive task.
Examples include estimating the covariance matrix of a clustering statistic
from a set of simulations and best-fit parameter estimation for a model.
 \nbkit{} implements a \code{TaskManager} utility to allow
users to easily iterate over multiple tasks while executing in parallel.
Users can specify the desired number of processes assigned to each task,
and the \code{TaskManager} will iterate through the tasks, ensuring that all
processes are being utilized.

\section{Capabilities}
\label{sec:capabilities}

In this section, we provide a more detailed overview of some of the main
components of \nbkit{}. In particular, we describe how cosmology
calculations are performed (\S\ref{sec:cosmo-details}), outline the available
Catalog (\S\ref{sec:catalog-details}) and Mesh (\S\ref{sec:mesh-details})
classes, and provide details and references for the various
algorithms currently implemented (\S\ref{sec:algorithm-details}).

\subsection{Cosmology}
\label{sec:cosmo-details}

The \code{nbodykit.cosmology} module includes functionality for
representing cosmological parameter sets and computing
various common theoretical quantities in LSS that depend on the background
cosmological model. The underlying engine for these calculations is
the \code{CLASS} Boltzmann solver \citep{Blas:2011,Lesgourgues:2011}.
We use the Python bindings of the \code{CLASS} C library provided
by the \classylss{} package. Comparing to the binding provided by
the \code{CLASS} source code, \classylss{} is a direct mapping of the
\code{CLASS} object model to Python and integrates with the
\numpy{} array protocol natively.

The main object in the module is the \code{Cosmology} class,
which users can initialize by specifying a unique set of cosmological
parameters (using the syntax of \code{CLASS}). This class represents
the background cosmological model and contains methods to compute
quantities that depend on the model. Most of the \code{CLASS} functionality
is available through methods of the \code{Cosmology} object.
Examples include distance as a function of redshift $z$, the Hubble parameter
$H(z)$, the linear power spectrum, the nonlinear power spectrum,
and the density and velocity transfer functions.
Several \code{Cosmology} objects are provided for
well-known parameter sets, including the \textit{WMAP} 5, 7, and 9-year
results \citep{Komatsu:2009,Komatsu:2011,Hinshaw:2013} and the
\textit{Planck} 2013 and 2015 results \citep{Planck:2014,Planck:2016}.

The \code{nbodykit.cosmology} module also includes classes to represent
theoretical power spectra and correlation functions. The \code{LinearPower}
class can compute the linear power spectrum as a function of redshift
and wavenumber, using either the transfer function as computed by
\code{CLASS} or the analytic approximations of \citet{Eisenstein:1998b}.
The latter includes the so-called ``no-wiggle'' transfer function,
which includes no BAO but the correct broadband features and is
useful for quantifying the significance of potential BAO features.
Similarly, we provide the \code{NonlinearPower} object to compute nonlinear
power spectra, using the \code{Halofit} implementation in \code{CLASS}
\citep{Smith:2003}, which includes corrections from \citet{Takahashi:2012}.
The \code{ZeldovichPower} class uses the linear power spectrum object to
compute the power spectrum in the Zel'dovich approximation (tree-level
Lagrangian perturbation theory). The implementation closely follows the
appendices of \citet{Vlah:2015} and relies on a Python implementation and
generalization of the
\code{FFTLog} algorithm\footnote{\url{https://github.com/eelregit/mcfit}}
\citep{Hamilton:2000}. Finally, we also provide
a \code{CorrelationFunction} object to compute theoretical
correlation functions from any of the available power classes
(using \code{FFTLog} to compute the Fourier transform).

We choose to use the \code{CLASS} software for the cosmological engine in
\nbkit{} rather than the most likely alternative, the \code{astropy.cosmology}
module. This allows \nbkit{} to leverage the full power of a Boltzmann solver
for LSS calculations. We provide syntax compatibility between the
\code{Cosmology} class and
\astropy{} when appropriate and provide functions to transform between the
cosmology classes used by the two packages. However, we note that there are
important differences between the two implementations. In particular, the
treatment of massive neutrinos differs, with \astropy{} using the
approximations of \citet{Komatsu:2011} rather than the direct calculations,
as in \code{CLASS}.

\subsection{Catalogs}
\label{sec:catalog-details}

In this section, we describe the two main ways that catalogs are created
in \nbkit{}, as well as tools for cleaning and manipulating data stored
in Catalog objects.

\subsubsection{Reading Data from Disk}

We provide support for loading data from disk into Catalog objects for
several of the most common data storage formats in LSS data analysis.
These formats include
plaintext comma-separated value (CSV) data
\citep[via \pandas{},][]{McKinney:2010},
binary data stored in a columnar format,
HDF5 data \citep[via \hdfpy{},][]{h5py},
FITS data \citep[via \fitsio{},][]{fitsio},
and the \bigfile{} data format.
We also provide more specialized readers for particle data from
the Tree-PM simulations of \citet{White:2002}
and the legacy binary format of the GADGET simulations \citep{Springel:2005}.
These Catalog objects use the \code{nbodykit.io} module, which includes
several ``file-like'' classes for reading data from disk. These
file-like objects implement a \code{read()} function
that provides the random-read interface which
returns a slice of the data for the requested columns.
Users can easily support custom file formats by implementing their own
subclass and \code{read()} interface.

Formats storing data on disk in a columnar format yield the best performance
results, as the entirety of the data does not need to be parsed to yield the
desired slice of the data on a given process.
This is not true for the CSV storage format. We mitigate performance
issues by implementing an enhanced version of the CSV
parser in \pandas{} that supports faster parallel random access.
Our preferred IO format, \bigfile{}, is massively parallel and stores
data via a columnar format.

Finally, the Catalog object supports loading data from multiple files at
once, providing a continuous view of the entirety of the data.
This becomes particularly powerful when combined with the random-read interface,
as arbitrary slices of the combined data can be accessed. For example, a single
Catalog object can provide access to arbitrary slices of the output binary
snapshots from an $N$-body simulation (stored over multiple files), often
totaling 10-100 GB in size.

\subsubsection{Generating Catalogs at Runtime}

\nbkit{} includes several Catalog classes that generate simulated
data at runtime. The simplest of these allows
generating random columns of data in parallel using the \code{numpy.random}
module. We also provide a \code{UniformCatalog} class that generates
uniformly distributed particles in a box. These classes are useful for testing
purposes, as well as for use as unclustered, synthetic data in clustering
estimators.

\nbkit{} also includes functionality for generating more
realistic approximations of large-scale structure. \code{LogNormalCatalog}
generates a set of objects by Poisson sampling a log-normal density field and
applies the Zel’dovich approximation to model nonlinear evolution
\citep{Cole:1991,Agrawal:2017}. The user can specify the input linear power
spectrum and the desired output redshift of the catalog.

Catalog objects can also be created using the mock generation techniques
of the \halotools{} software \citep{Hearin:2017} for populating halos with
objects. \halotools{} includes functionality for populating halos via a wide
range of techniques, including the halo occupation distribution (HOD),
conditional luminosity function, and abundance matching methods. We refer the
reader to \citet{Hearin:2017} for further details. \nbkit{} supports
using a generic \halotools{} model to populate a halo catalog. We also
include built-in, specialized support for the HOD models of
\citet{Zheng:2007}, \citet{Leauthaud:2011}, and \citet{Hearin:2016}.

Finally, the \fastpm{} package implements an \nbkit{} Catalog object
that generates particles via the FastPM approximate $N$-body simulation
scheme \citep{Feng:2016b}. The FastPM library is
massively parallel and exhibits excellent strong scaling with the number of
available processes (see \S\ref{sec:benchmarks}).

\subsubsection{On-demand Data Cleaning}

\nbkit{} uses the \dask{} library \citep{dask}
to represent the data columns of a Catalog object as \dask{} array objects
instead of using the more familiar \numpy{} array. The \dask{} array has two
key features that help users work interactively with data, and, in particular,
large data sets. The first feature is \textit{delayed evaluation}.
When manipulating a \dask{} array, operations are not evaluated
immediately but instead stored in a task graph.
Users can explicitly evaluate the \dask{} array (returning a \numpy{}
array) via a call to a \code{compute()} function.
Second, \dask{} arrays are \textit{chunked}. The array object is internally
divided into many smaller arrays, and calculations are performed on
these smaller ``chunks.''

The delayed evaluation of \dask{} arrays is particularly useful during
the process of data cleaning, when users manipulate input data before
feeding it into the analysis pipeline. Common examples
of data cleaning include changing
the coordinate system from galactic to Euclidean, converting between
unit conventions, and applying masks. When using large data sets,
the time to load the full data set into memory can be significant.
This delay hinders data exploration and limits the interactive benefits
of the Python language. \dask{} arrays
allow users to design data-cleaning pipelines on the fly.
If the data format on disk supports random-read access,
users can easily select and peek at a small subset of data without
reading the full data set. This becomes especially useful when prototyping
scientific models in an interactive environment, such as a \jupyter{} notebook.

The chunked nature of the \dask{} array allows array
computations to be performed on large data sets that do not fit into memory
because the chunk size defines the amount of data loaded into memory at
any given time. It effectively extends the maximum size of
useable data sets from the size of memory to the size of the disk storage.
This feature also simplifies the process of dealing
with large data sets in interactive environments.

\subsection{Meshes}
\label{sec:mesh-details}

\subsubsection{Painting a Mesh}

The Mesh object implements a \code{paint()} function, which is
responsible for generating the field values on the mesh and returning an
array-like object to the user. Meshes provide an equal treatment of
configuration and Fourier space, and users can specify whether the painted
array is defined in configuration or Fourier space.
In the former case, a \code{RealField} is returned and in the latter, a
\code{ComplexField}. These objects are implemented by the \pmesh{}
package and are subclasses of the \numpy{}
\code{ndarray} class. They are related via a
real-to-complex parallel FFT operation, implemented using \pfftpython{} via
the \code{r2c()} and \code{c2r()} functions.

The \code{paint()} function paints mass-weighted (or equivalently,
number-weighted) quantities to the mesh. The field that is painted is

\begin{equation}
   F(\vx) = \left[ 1 + \delta'(\vx) \right] V(\vx),
\end{equation}
where $V(\vx)$ represents the field value painted to the mesh and
$\delta'(\vx) =  n'(\vx)/\bar{n}' - 1$ is the weighted overdensity field.
It is related to the unweighted number density as $n'(\vx) = W(\vx) n(\vx)$,
where $W(\vx)$ are the weights.

In \nbkit{}, users can control the
behavior of both $V(\vx)$ and $W(\vx)$. In the default case, both quantities
are unity, and the field painted to the mesh is $1 + \delta$. As an
illustration, $V(\vx)$ can be specified as a velocity component to paint the
momentum field (mass-weighted velocity).
We also provide a mechanism by which users can further transform the painted
field on the mesh. The \code{apply()} function can be used to apply
a function to the mesh, either in configuration or Fourier space. Multiple
functions can be applied to the mesh, and the operations are performed when
\code{paint()} is called.

\subsubsection{From Catalog to Mesh}

All Catalog objects include a \code{to\_mesh()} function which creates a
Mesh object using the specified number of cells per mesh side. This function
allows users to configure exactly how the catalog is interpolated
onto the mesh. Users can choose from several different mass assignment
windows, including the Cloud-In-Cell (CIC), Triangular Shaped Cloud (TSC),
and Piecewise Cubic Spline (PCS) schemes
\citep{Hockney:1981}. The Daubechies wavelet \citep{Daubechies:1992} and its
symmetric counterpart (``Symlets'', see, e.g.,
\code{PyWavelets}\footnote{\url{https://pywavelets.readthedocs.io}})
are also available. By default, the CIC window is
used. The interlacing technique \citep{Hockney:1981,Sefusatti:2016}
can reduce the effects of aliasing in Fourier space. In this scheme, the
Catalog object is interpolated onto
two separate meshes separated by half of a cell size. When the fields are
combined in Fourier space, the leading-order
contribution to aliasing is eliminated.

Users can also configure whether or not the window is \textit{compensated},
which divides the density field in Fourier space by \citep{Hockney:1981}

\begin{equation}\label{eq:window-deconvolution}
W(\vk) = \Pi_i \left[\mathrm{sinc}\left(\pi k_i/2\knyq\right) \right]^p,
\end{equation}
where $i \in \{x,y,z\}$, $p=2,3,4$ for CIC, TSC, and PCS, respectively,
and $\mathrm{sinc}(x) \equiv \sin(x)/x$. The Nyquist frequency of the
mesh is given by $\knyq = \pi N / L$, where $L$ is the
box size, and $N$ is the number of cells per box side.

\begin{figure}[tb]
\center
\includegraphics[scale=0.8]{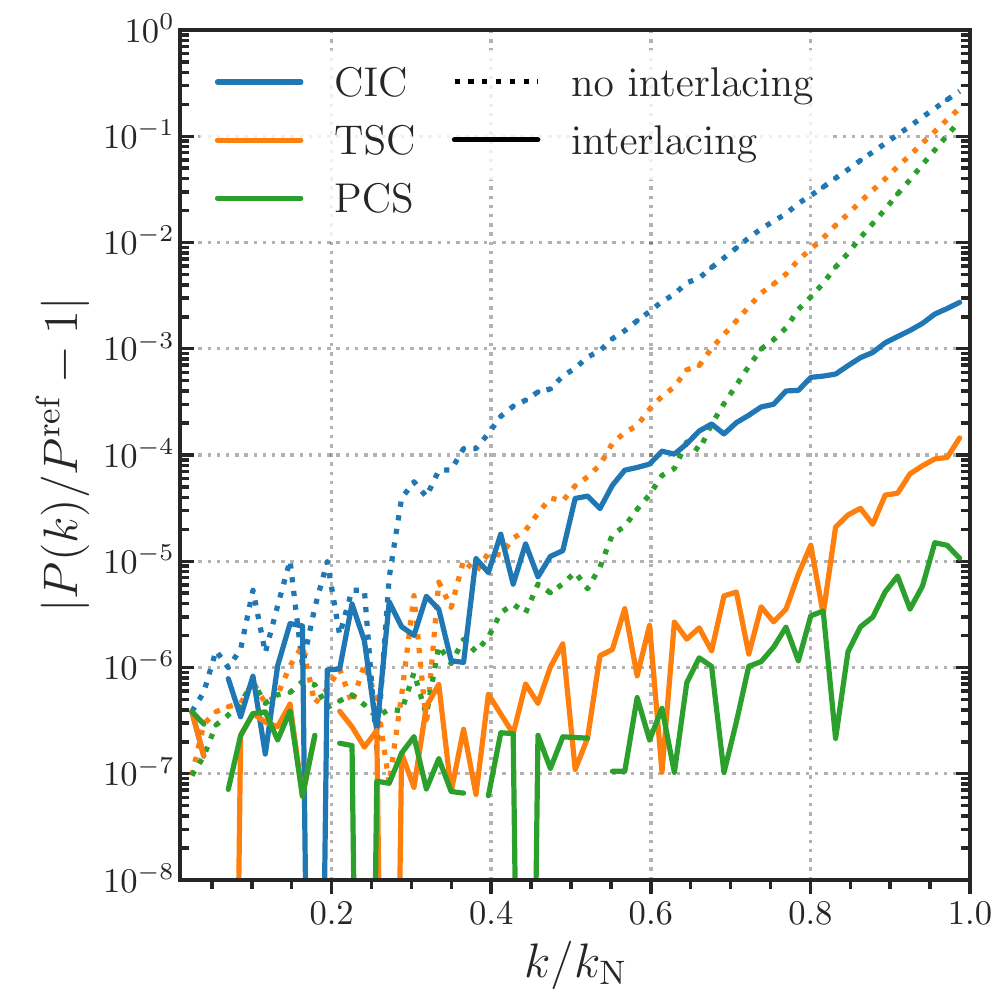}
\caption
{
A comparison of the effects of interlacing when using the CIC, TSC, and PCS
windows. We show the ratio of the power spectrum computed for a
log-normal density field using a mesh with $512^3$ cells to a
reference power spectrum $P^\mathrm{ref}$, computed using a mesh with
$1024^3$ cells. The ratio is shown as a function of
wavenumber in units of the Nyquist frequency of the lower-resolution mesh.
In all cases, the appropriate window compensation is performed using
equation~\ref{eq:window-deconvolution}.
}
\label{fig:window-corrs}
\end{figure}

\begin{figure}[tb]
\center
\includegraphics[width=\columnwidth]{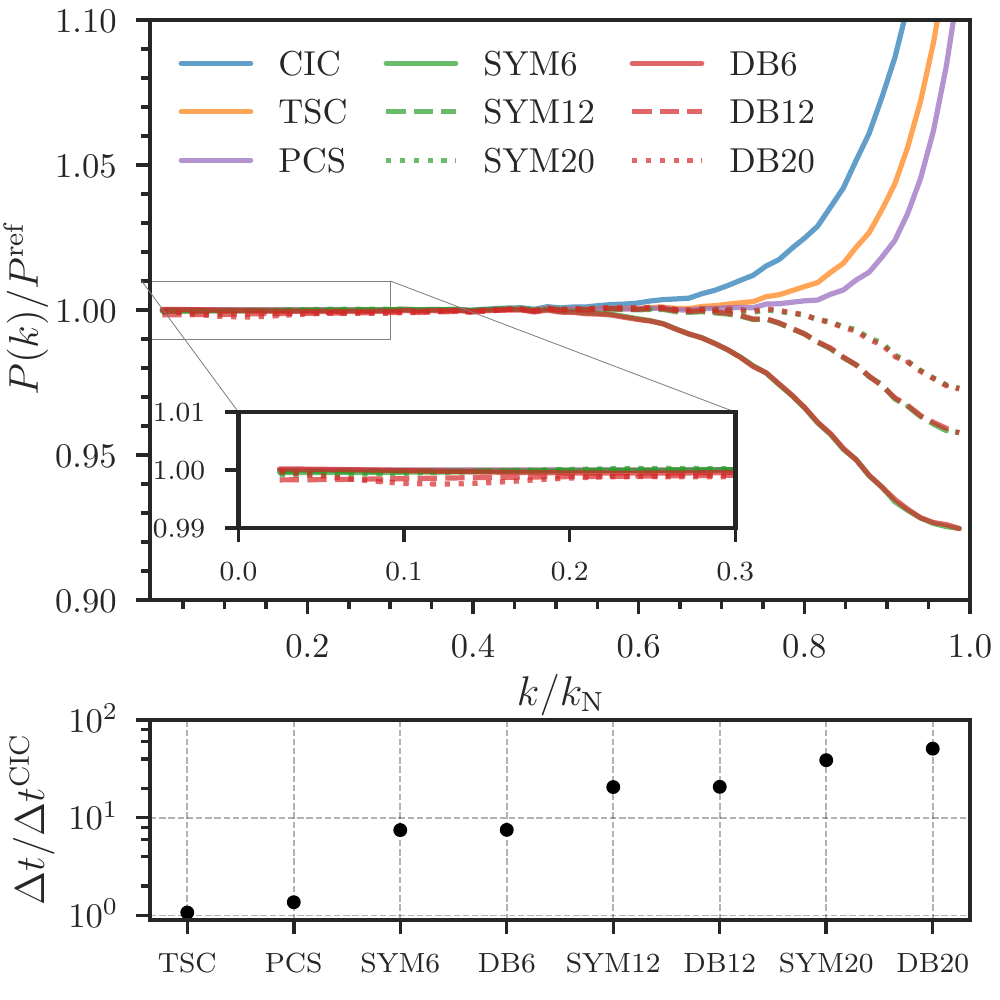}
\caption
{
The performance of the Daubechies
and Symlet wavelets in comparison to the CIC, TSC,
and PCS windows. Wavelet windows of sizes $a=$ 6, 12, and 20 are shown.
\textit{Top}: the ratio of the measured power to the reference
power spectrum, as in Figure~\ref{fig:window-corrs}.
Here, we apply no corrections when using the wavelet
windows and apply equation~\ref{eq:window-deconvolution}
for the CIC, TSC, and PCS windows. No interlacing is used for this test.
\textit{Bottom}: the speed of each interpolation window, relative to the
CIC window. Speeds were recorded when computing the power spectra
in the top panel.
}
\label{fig:wavelets}
\end{figure}

We provide comparisons of the various interpolation windows and correction
methods in this section. First, Figure~\ref{fig:window-corrs} illustrates the
effects of interlacing when using the CIC, TSC, and PCS schemes. This
comparison is similar to the detailed
analysis presented in \citet{Sefusatti:2016}. Second, we show the
effectiveness of the wavelet windows at reducing aliasing in
Figure~\ref{fig:wavelets}. For both figures, we paint a
\code{LogNormalCatalog} of $5 \times 10^7$ objects to a mesh of $512^3$ cells
in a box of side length $2500 \hMpc$. We compare the
measured power spectrum to a ``reference'' power spectrum, computed
using a mesh of $1024^3$ cells and the PCS window.
When using the CIC, TSC, and PCS windows, we de-convolve the
interpolation window using equation~\ref{eq:window-deconvolution}, while
we apply no such corrections when using wavelet-based windows.

Figure~\ref{fig:window-corrs} confirms the results of
\citet{Sefusatti:2016}---the interlacing technique performs very well at
reducing the effects of aliasing on the measured power spectrum. We achieve
sub-percent accuracy up to the Nyquist frequency when combining interlacing
with the CIC, TSC, and PCS windows. In general, higher-order windows perform
better, with the PCS scheme achieving a precision of at least
$\sim$$10^{-5}$ up to the Nyquist frequency.

Figure~\ref{fig:wavelets} compares the performance of the
Daubechies and Symlet wavelets to the CIC, TSC, and PCS windows.
As in Figure~\ref{fig:window-corrs},
we plot the ratio of the power spectrum computed using meshes of size
$512^3$ and $1024^3$ cells. We apply equation~\ref{eq:window-deconvolution}
for the CIC, TSC, and PCS windows but do not apply any corrections when
using the wavelet windows. For this comparison, we do not use interlacing.
We are able to confirm the results of \citet{Cui:2008} and
\citet{Yang:2009}, which claim 2\% accuracy on the power spectrum up to
$k\approx 0.7k_\mathrm{N}$ when using the DB6 window without any additional
corrections. However, the wavelet windows fail to match the precision achieved when using
interlacing, even when using the largest wavelet size tested here ($a=20$).
Furthermore, the Daubechies windows
introduce scale-dependence on large scales due to symmetry breaking
(see the inset of Figure~\ref{fig:wavelets}). The symmetric Symlet wavelets
do not suffer from this issue but also cannot match the accuracy
achieved when using interlacing.

Figure~\ref{fig:wavelets} also displays the relative speeds of each
of the windows discussed in this section (bottom panel). These timing
tests were performed
using 64 processes on the NERSC Cori Phase I system. The wavelet windows are
all significantly slower than the CIC, TSC, and PCS windows.
The TSC and PCS methods are only marginally slower than the
default CIC scheme, with slowdowns of $\sim$10\%
and $\sim$40\%, respectively. The CIC, TSC, and PCS windows rely on
optimized implementations in \pmesh{}, while the wavelet windows
use a slower lookup table implementation. Due to the precision of the
interlacing technique and the relative speed of the TSC and PCS windows,
we recommend using these options in most instances. However, it is generally
best to determine the optimal set of parameters for a particular
application by running convergence tests with different parameter
configurations.

\subsubsection{An Illustrative Example}

\begin{figure}[tb]
\begin{minipage}[c]{0.46\textwidth}
\center
\subfloat{\fbox{\includegraphics[width=\columnwidth]{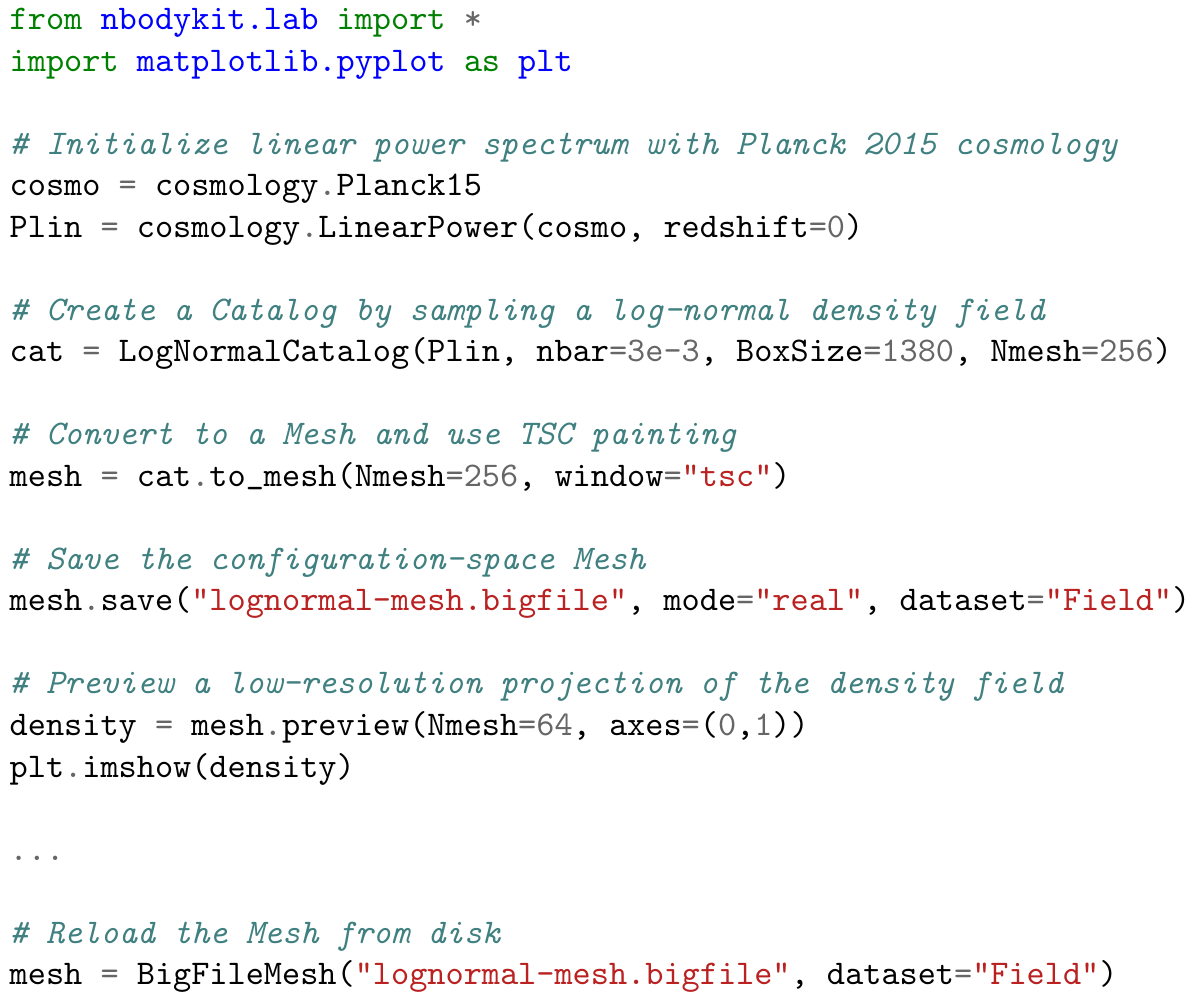}}}
\end{minipage}%
\begin{minipage}[c]{0.5\textwidth}
\center
\subfloat{\includegraphics[width=0.9\columnwidth]{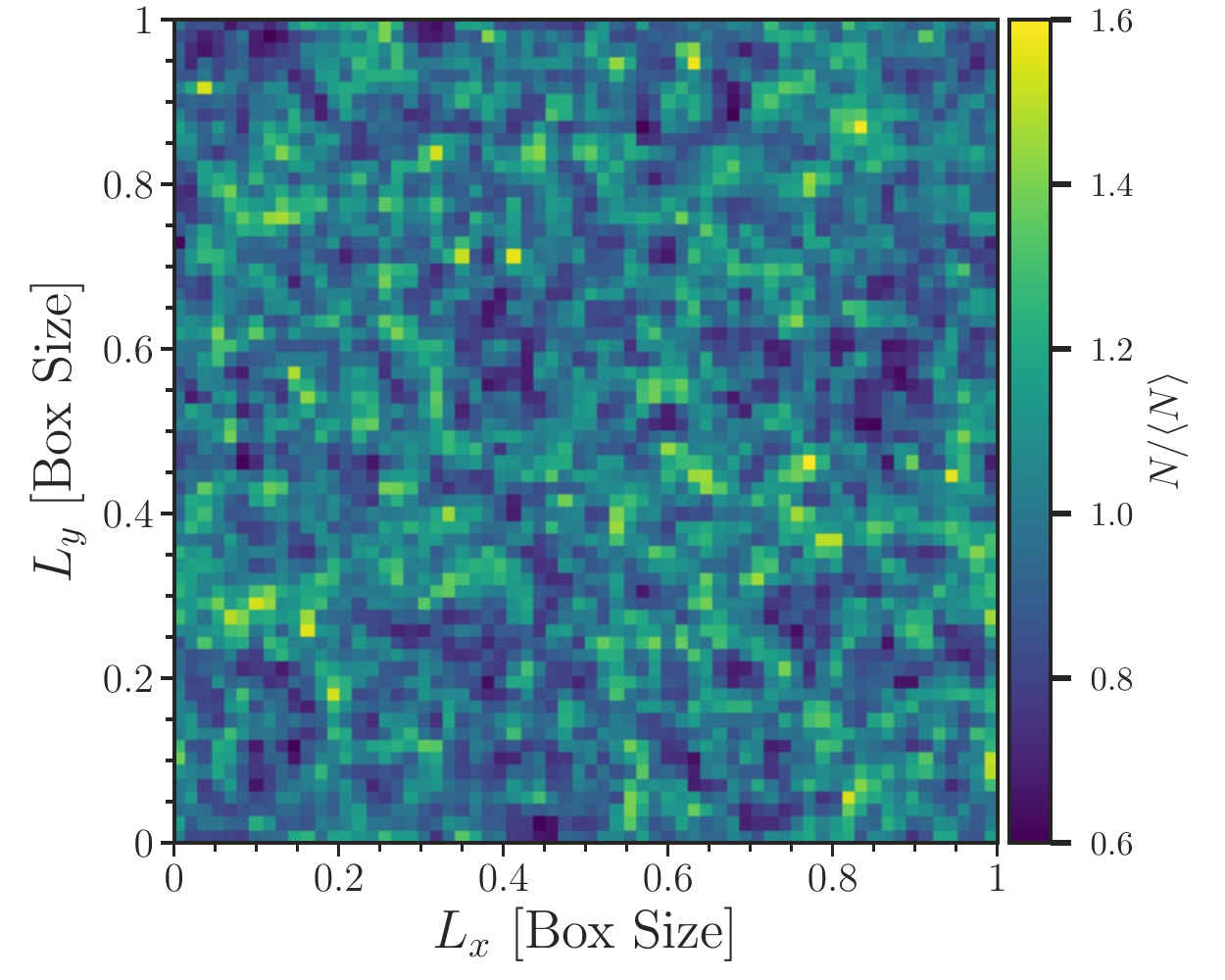}}
\end{minipage}
\caption{\textit{Top}: an analysis pipeline illustrating the creation of a
Mesh object from a Catalog, as well as how to serialize the painted mesh to
disk and preview a low-resolution projection of the density field for inspection.
\textit{Bottom}: the two-dimensional, low-resolution preview of the painted
density field $N/\langle N \rangle = 1 + \delta$.}
\label{fig:mesh-preview}
\end{figure}

We demonstrate the use of Mesh objects by example in
Figure~\ref{fig:mesh-preview}, which gives a short code snippet that
creates a Mesh object from an existing Catalog,
saves the configuration space density field to disk, and then reloads
the data into memory. The snippet also demonstrates the \code{preview()}
function, which can create a lower resolution projection of the full mesh
field. This can be useful to quickly inspect mesh fields interactively, which
would otherwise be difficult due to memory limitations. We show the
preview of the density field from a log-normal catalog in the bottom
panel of Figure~\ref{fig:mesh-preview}, where the large-scale
structure is clearly evident, even in the low-resolution projection.

\subsection{Algorithms}
\label{sec:algorithm-details}

The \code{nbodykit.algorithms} module includes parallel
implementations of some of the most commonly used large-scale structure
analysis algorithms. We take care to provide support for data sets from
both observational
surveys and $N$-body simulations. In this section, we provide an
overview of the available functionality.
The set of algorithms currently implemented is not meant to be
exhaustive, but instead a solid foundation for LSS data analysis.

\subsubsection{Power Spectra}

For simulation boxes with periodic boundary conditions,
the \code{FFTPower} algorithm measures the power directly from
the square of the Fourier modes of the overdensity field. The 1D or 2D
power spectrum, $P(k)$ or $P(k,\mu)$, can be computed, as well as the
power spectrum multipoles $P_\ell(k)$. Here, $\mu$ represents the
angle cosine between the pair separation vector and the line-of-sight.
For observational data, in the
form of right ascension (RA), declination (Dec), and redshift, the power spectrum
multipoles of the density field can be computed using the
\code{ConvolvedFFTPower} algorithm. The implementation uses
the FFT-based estimator described in \citet{Hand:2017}, which requires
$2\ell+1$ FFTs to compute a given multipole of order $\ell$. This estimator
improves the FFT-based estimator presented by
\citet{Bianchi:2015} and \citet{Scoccimarro:2015}, building on the ideas of
previous power spectrum estimators \citep{Feldman:1994,Yamamoto:2006},
and in particular, the treatment of the anisotropic 2PCF using
spherical harmonics of \citet{Slepian:2015}.
We also provide the \code{ProjectedFFTPower} for computing
the power spectrum of a field in a simulation box, projected along the
specified axes. Such an observable is useful for e.g., Lyman-$\alpha$ or
weak lensing data analysis. The correctness of these algorithms has been
verified using independent implementations
from within the Baryon Oscillation Spectroscopic Survey (BOSS) collaboration.

\subsubsection{2-Point Correlation Functions}

\nbkit{} includes functionality for counting pairs of objects and computing
their correlation function in configuration space. We leverage the
blazing speed of the publicly available \Corrfunc{} chaining mesh code for
these calculations \citep{Sinha:2017}. We adapt its
highly optimized pair counting routines to perform calculations using MPI.
We perform a domain decomposition on the input data such that the objects
on a particular MPI rank are spatially confined to include all pairs within
the maximum separation. For non-uniform density fields,
the domain decomposition results in a particle load that
is balanced across MPI ranks.\footnote{We thank Biwei Dai for the implementation of the load balancer.}
The relevant pair counting algorithms
are \code{SimulationBoxPairCount} and \code{SurveyDataPairCount}. These
classes can count pairs of objects as a function of the 3D separation $r$,
the separation $r$ and angle to the line-of-sight $\mu$, the angular
separation $\theta$, and the projected distances perpendicular $r_p$ and
parallel $\pi$ to the line-of-sight.

Users can compute the correlation function of a Catalog using the
\code{SimulationBox2PCF} and \code{SurveyData2PCF} classes, which internally
rely on the previously described pair counting classes. For data
with periodic boundary conditions, we use analytic randoms to estimate
the correlation function using the so-called ``natural'' estimator:
$DD/RR - 1$. A Catalog object holding synthetic randoms can be supplied,
in which case the Landy-Szalay estimator \citep{Landy:1993}
is employed: $(DD - 2 DR + RR)/RR$. The variations of the correlation
function that can be computed by these two classes are as follows:
\begin{itemize}
  \item as a function of three-dimensional separation, $\xi(r)$
  \item accounting for the angle to the line-of-sight, $\xi(r, \mu)$ and $\xi(r_p, \pi)$
  \item as a function of angular separation, $w(\theta)$
  \item projected over the line-of-sight separations,  $w_p(r_p)$
\end{itemize}
The correctness of the pair counting and correlation function algorithms
described here was independently verified using the \kdcount{} and
\halotools{} software.

\subsubsection{3-Point Correlation Function}

The \code{SimulationBox3PCF} and \code{SurveyData3PCF} classes compute
the multipoles of the isotropic 3-point correlation function (3PCF) in
configuration space. The algorithm follows the implementation described in
\citet{Slepian:2015}, which scales as $\O(N^2)$, where $N$ is the
number of objects. Their improved estimator relies on a spherical harmonic
decomposition to achieve a similar scaling with $N$ as two-point clustering
estimators. We note that the FFT-based implementation of
this algorithm \citep[presented in][]{Slepian:2016} and the anisotropic
version described in \citet{Slepian:2017b} have not yet been implemented,
although there are plans to do so in the future. We have verified the
accuracy of the isotropic 3PCF classes against the implementation
used in \citet{Slepian:2015}. An implementation
of this algorithm including anisotropy written in C++ and
optimized for HPC machines was recently
presented in \citet{Friesen:2017}.

\subsubsection{Grouping Methods}

The \code{FOF} class implements the well-known Friends-of-Friends algorithm,
which identifies clusters of points that are spatially
less distant than a threshold linking length. It uses a
parallel implementation of the algorithm described in \citet{Feng:2017},
which utilizes KD-trees and the \kdcount{} software.
FOF groups can be identified as a function of
three-dimensional or angular separation.
We also provide functions for transforming the output of the \code{FOF}
algorithm to a Catalog of halo objects (a \code{HaloCatalog})
in a manner compatible with the \halotools{} software.

\nbkit{} can also identify clusters of objects using
a cylindrical rather than spherical geometry. We implement a parallel
version of the algorithm described in \citet{Okumura:2017} in the
\code{CylindricalGroups} class. Our implementation relies on
the neighbor querying capability of \kdcount{} and the group-by methods
of \pandas{}.

Finally, the \code{FiberCollisions} class simulates the process of assigning
spectroscopic fibers to objects in a fiber-fed redshift survey
such as BOSS or eBOSS \citep{Dawson:2013, Dawson:2016}. This procedure
results in so-called ``fiber collisions'' when two objects are separated
by an angular width on the sky that is smaller than the fiber size. We
follow the procedure outlined in \citet{Guo:2012} to assign fibers to an input
catalog of objects. We identify angular FOF groups using a linking
length equal to the fiber collision scale and assign fibers to the objects
in such a manner as to minimize the number of objects that do not receive a
fiber.

\subsubsection{Miscellaneous}

\nbkit{} also includes algorithms that generally serve a supporting role
in other algorithms. The \code{KDDensity} class estimates a proxy density
quantity for an input set of objects using the inverse cube of the distance to
an object's nearest neighbor. The \code{RedshiftHistogram} class computes the
mean number density as a function of redshift, $n(z)$, from an input catalog of
objects. We plan to generalize this algorithm to be a more universal histogram
calculator that could, for example, compute mass or luminosity functions.

\section{Development Workflow}
\label{sec:dev-approach}

\subsection{Version Control}

\nbkit{} is developed using the version control features of
\texttt{git},\footnote{\url{http://git-scm.com}} and the code is hosted
in a public repository on
GitHub.\footnote{\url{http://github.com/bccp/nbodykit}}
Major changes to the code base are
performed using a \textit{pull request} workflow, which provides a
mechanism for developers to review changes before they are merged into the
main source code. Users can contribute to \nbkit{} by first \textit{forking}
the main repository, making changes in this fork and submitting the
changes to the main repository via a pull request. This workflow helps
assure the overall quality of the code base and ensures that new changes
are properly documented and tested. Bugs and new feature requests can be
submitted as GitHub issues. Alternatively, users can send an email
to \url{nbodykit-issues@fire.fundersclub.com}, which will automatically
open an issue on GitHub. As \nbkit{} is intended as a community-based
resource, we encourage user contributions and ideas for new functionality.
We adopt a ``mentoring'' approach for new features and will gladly offer
advice and guidance to new users who wish to contribute to \nbkit{} for the
first time.

\subsection{Automated Testing with MPI Support}

\nbkit{} is extensively tested via hundreds of unit tests using the
\runtests{}\footnote{\url{https://github.com/rainwoodman/runtests}} package
\citep{runtests}. As \mpipy{} does not
provide a reusable framework for testing parallel applications, we have
developed \runtests{} to fill this gap in the development process.
It extends the \code{py.test}\footnote{\url{http://pytest.org}}
testing framework, adding several features. First, the test driver
incrementally rebuilds and installs the Python package before running the
test suite. Second, it adds MPI support by allowing users to specify the
number of processes with which each test function should be executed.
It also supports computing the testing coverage for parallel applications,
where test coverage is defined as the percentage of the software
covered by the test suite.

We execute the \nbkit{} test suite via the continuous integration (CI) service
Travis,\footnote{\url{https://travis-ci.org}} using \runtests{} to test
both serial and parallel execution of the code. The test suite is
currently executed on both Linux and Mac OS X operating systems and for Python
versions 2.7, 3.5, and 3.6. Whenever a pull request is opened, the test suite
is executed and the new changes will not be merged if the test suite fails.
We also compute the testing coverage of the code base. Currently, \nbkit{}
maintains a value of 95\%. We use the
Coveralls\footnote{\url{https://coveralls.io}} service to ensure that new
changes cannot be merged into the main repository if the testing coverage
decreases.

\subsection{Use on Personal and HPC Machines}\label{sec:installing}

\nbkit{} is compatible with both Python versions 2.7 and 3.x.
For personal computing systems (Mac OS X and Linux), we provide binaries of
\nbkit{} and its dependencies on the Berkeley Center for
Cosmological Physics (BCCP)
Anaconda channel.\footnote{\url{https://anaconda.org/bccp}}
\nbkit{} (and all of its dependencies) can be installed into an Anaconda
environment using a simple command: \texttt{conda install -c bccp nbodykit}.
We ensure all packages on the BCCP channel are up-to-date using a
nightly cron job hosted on the Travis CI service.

Supercomputing systems often require recompiling the
dependencies of \nbkit{} using the machine-specific compilers and MPI
configuration. For example, we use the ``conda build'' functionality of the
Anaconda package to compile and update \nbkit{} and its dependencies nightly
on the NERSC Cray supercomputers. The infrastructure for building
\nbkit{} and its dependencies is publicly available on
GitHub,\footnote{\url{https://github.com/bccp/conda-channel-bccp}} which users can
re-use to setup \nbkit{} on HPC machines other than NERSC. However,
we recommend that users first test if the default binaries on the BCCP
channel are compatible with their supercomputing environment.

The remaining barrier to using \nbkit{} on HPC systems is the incompatibility
of the Python launch system and the shared file systems of HPC machines.
When launching an MPI application using Python, the file system will stall
when all of the Python instances (can be thousands or more) query the
file system for modules on the search path. This issue effectively prevents
the use of Python applications on HPC machines.

\nbkit{} utilizes an open-source solution, denoted
``python-mpi-bcast'',
to facilitate deploying Python applications on HPC machines \citep{Feng:2016}.
This tool bundles and delivers
runtime dependencies to the HPC computing nodes via an MPI broadcast operation,
bypassing the file system bottleneck and allowing Python applications to launch
at near-native speed. Users can modify their job scripts
in a non-invasive manner to deploy our tool.
Additional details and setup instructions can be found in \citet{Feng:2016}.
The tool is publicly available on
GitHub.\footnote{\url{https://github.com/rainwoodman/python-mpi-bcast}}

\subsection{Documentation}

Documentation for \nbkit{} is available on
Read the Docs.\footnote{\url{http://nbodykit.readthedocs.io}} The documentation
is generated using \code{Sphinx}\footnote{\url{http://www.sphinx-doc.org}}
and includes comprehensive documentation of the \nbkit{} API. It also includes
detailed walk-throughs of each of the main components of \nbkit{}.

We provide a set of recipes detailing a broad selection of the
functionality available in \nbkit{} in the ``Cookbook'' section of the
documentation. Ranging from simple tasks to more
complex work flows, we hope that these recipes help users become acclimated
to \nbkit{} as well as illustrate the power of \nbkit{} for LSS data analysis.
The recipes are in the form of Jupyter notebooks. An interactive environment
containing the recipe notebooks is available
to users via the Binder service.\footnote{\url{https://mybinder.org}}
This allows new users to explore \nbkit{} without
installing \nbkit{} on their own machine.

\section{In Action}
\label{sec:inaction}

\begin{figure*}[tb]
  \begin{minipage}[c]{0.43\textwidth}
  \center
  \subfloat{\fbox{\includegraphics[width=\textwidth]{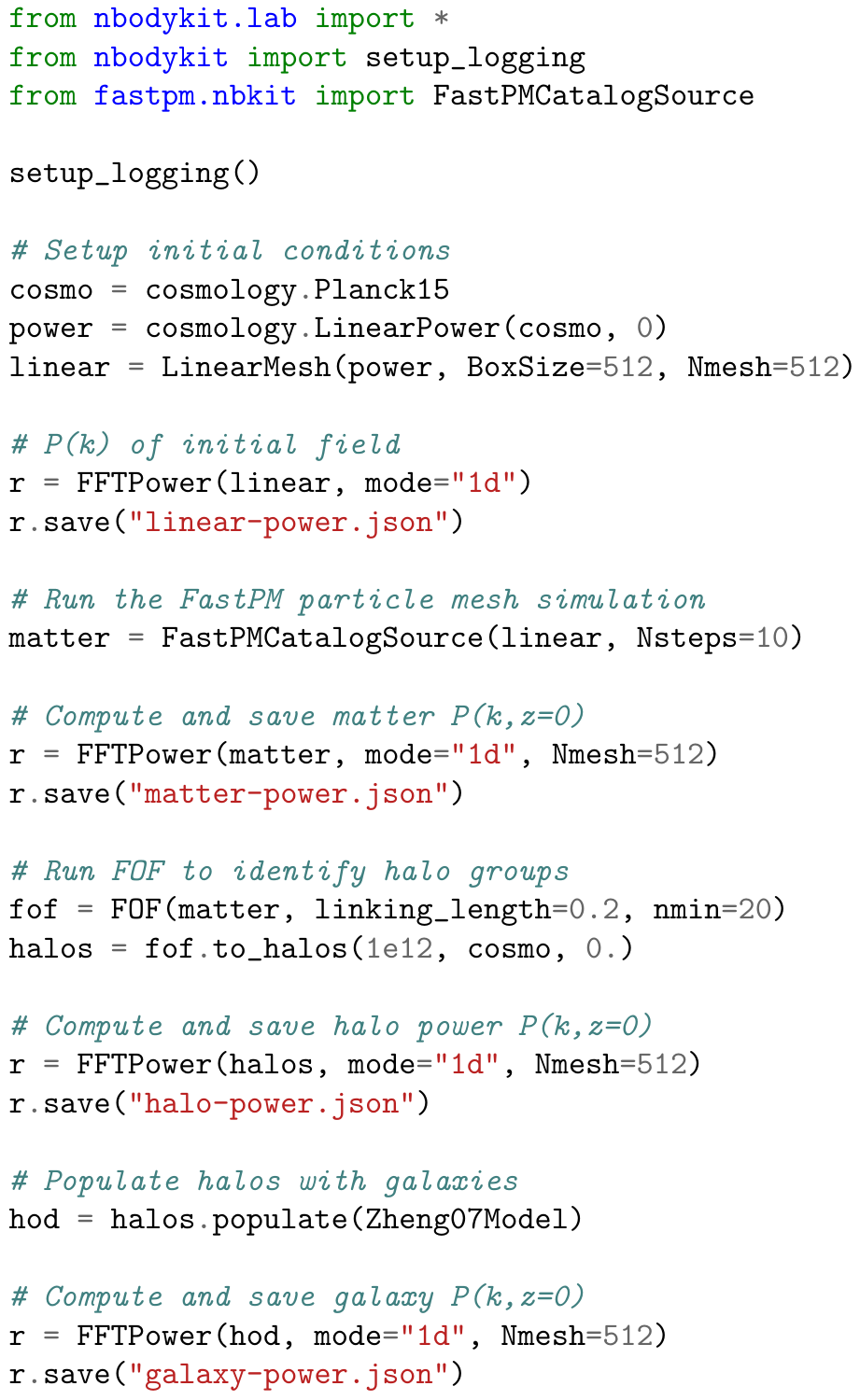}}}
  \end{minipage}%
  \begin{minipage}[c]{0.57\textwidth}
  \center
  \begin{minipage}[c]{\textwidth}
    \center
  \subfloat{\includegraphics[width=\textwidth]{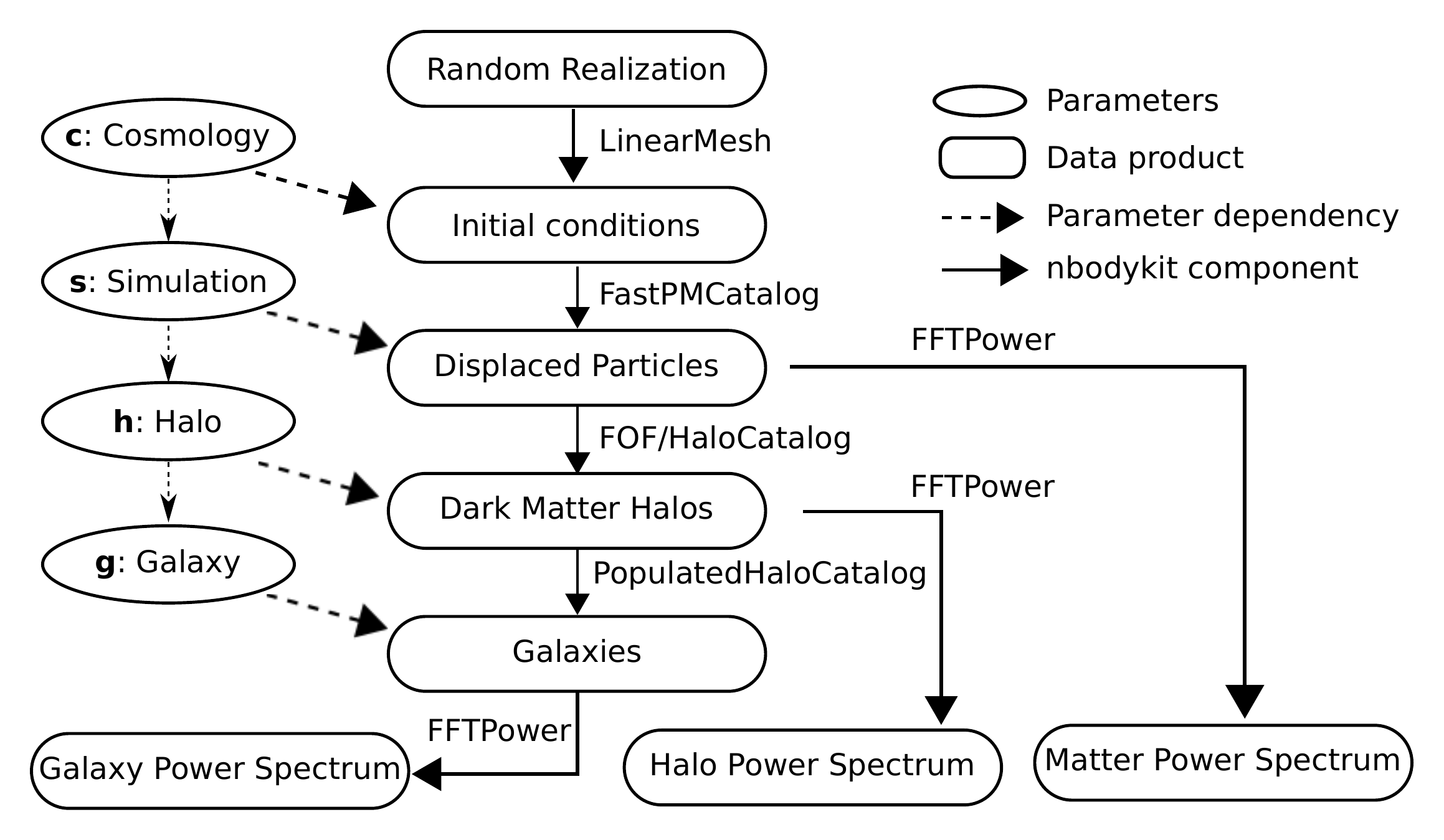}}
  \end{minipage}
  \begin{minipage}[c]{0.7\textwidth}
  \subfloat{\includegraphics[width=\textwidth]{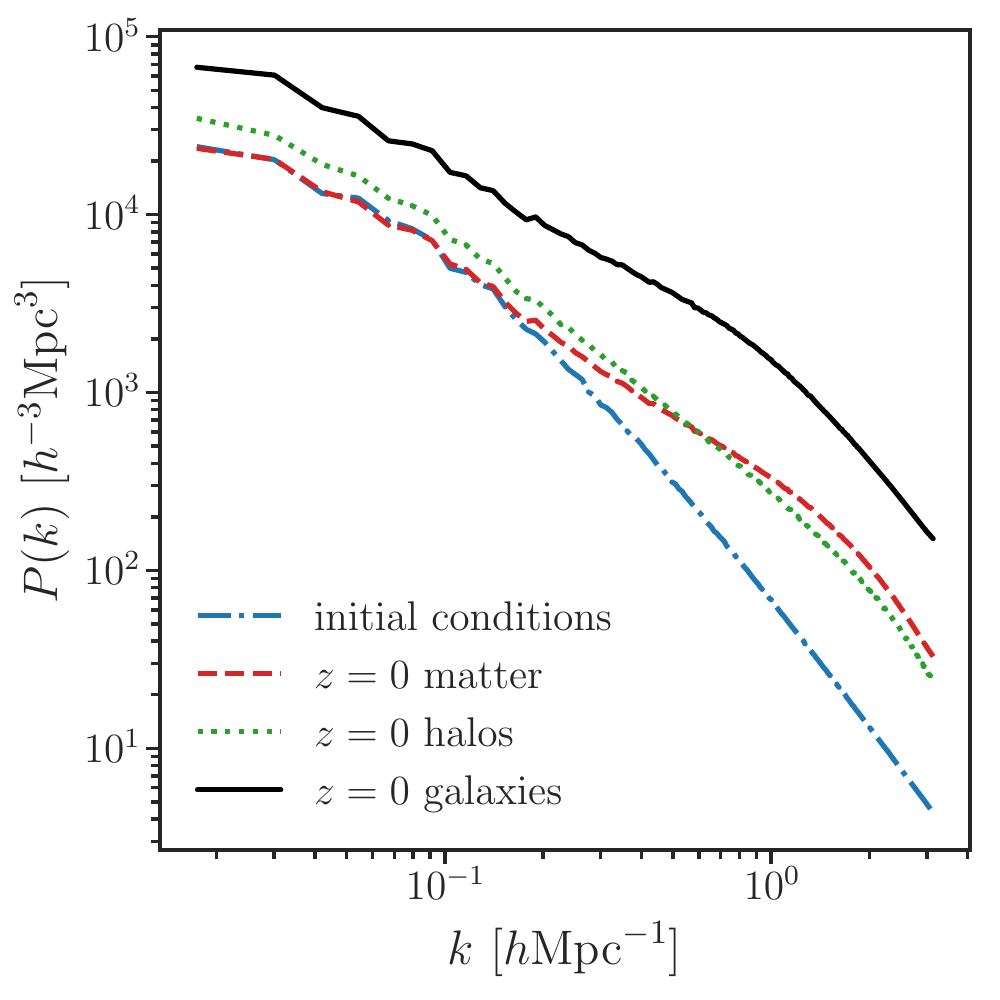}}
  \end{minipage}
  \end{minipage}%
\caption{A galaxy clustering emulator, implemented with \nbkit{}.
\textit{Left}: the source code for the application, which evolves an initial
Gaussian field to $z=0$ using the FastPM simulation scheme, identifies FOF halos,
populates those halos with galaxies, and records the power spectrum of each
step. \textit{Right, top}: the flow of data through the various components.
\textit{Right, bottom}: the resulting $P(k)$ measured for each
step in the emulator. Performance benchmarks for this application are given in
Figure~\ref{fig:emu-benchmark}.}
\label{fig:emulator-code}
\end{figure*}

In this section, we describe a realistic LSS application using \nbkit{}:
a galaxy clustering emulator. The goal of the emulator is
to produce the galaxy power spectrum
from first principles, given a background cosmological model.
The application combines several components of \nbkit{} to achieve
this goal. The steps include:

\begin{itemize}
  \item Initial conditions: the \code{LinearMesh} class creates a
  Gaussian realization of a density field in Fourier space from an input
  power spectrum.
  \item $N$-body simulation: the initial conditions are evolved
  forward to $z=0$ using the FastPM quasi-$N$-body particle
  mesh scheme of \citet{Feng:2016b}.
  \item Halo Identification: halos are identified from the matter
  field using the \code{FOF} grouping algorithm.
  \item Halo Population: halos are populated with galaxies using
  the HOD from \citet{Zheng:2007} and the \halotools{} package.
  \item Clustering Estimation: $P(k)$ is computed for each of the
  above steps using the \code{FFTPower} algorithm.
\end{itemize}

We diagram the flow of data and parameters for these steps in
the top right panel of Figure~\ref{fig:emulator-code}. We also show the
source code for the application using \nbkit{}, which can be
implemented using only $\sim$30 lines of code.
We emphasize that with the component-based approach of \nbkit{}, the user
is free to output and serialize any intermediate data products during the
execution of the larger application, as we have done in this example for the
power spectra of the initial, matter, and halo density fields.
Finally, note that the source code in Figure~\ref{fig:emulator-code} can be executed
with an arbitrary number of MPI ranks. We discuss performance benchmarks for
this application as a function of the number of MPI processes
in the next section.

\section{Performance Benchmarks}
\label{sec:benchmarks}

\begin{figure}[tb]
\center
\includegraphics[width=\columnwidth]{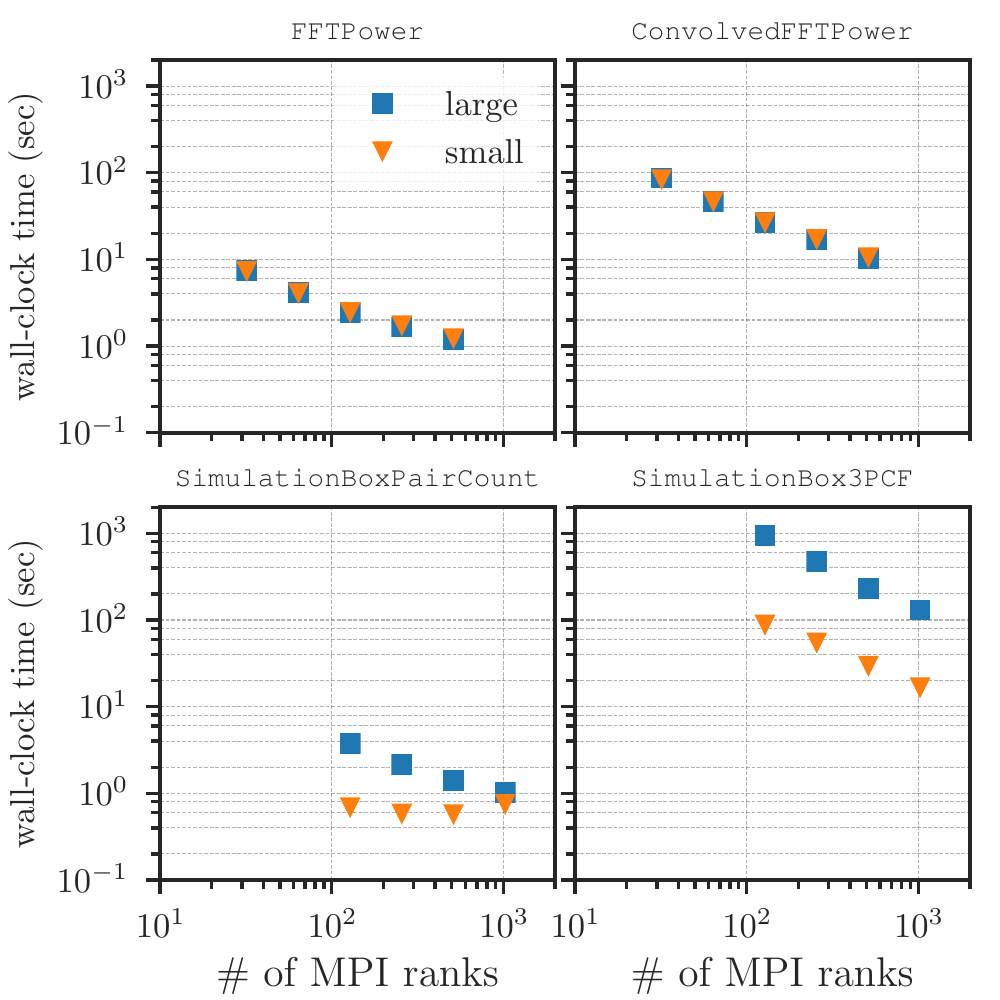}
\caption{Performance benchmarks for four \nbkit{} algorithms for our
``small'' data set ($10^6$ objects) and our ``large'' data set ($10^7$ objects).
The algorithms in the top row use FFT-based estimators to compute power spectra,
while those in the bottom row of panels count pairs of objects in
configuration space. The FFT-based algorithms take near-identical time for
the large and small data sets due to the use of a $1024^3$ mesh in both cases.
The benchmarks were performed on the NERSC Cori
Phase I Haswell nodes using 32 MPI ranks per node. See the text of
Section~\ref{sec:benchmarks} for further details on the test configurations.}
\label{fig:benchmarks}
\end{figure}

\begin{figure}[tb]
\center
\includegraphics[width=0.8\columnwidth]{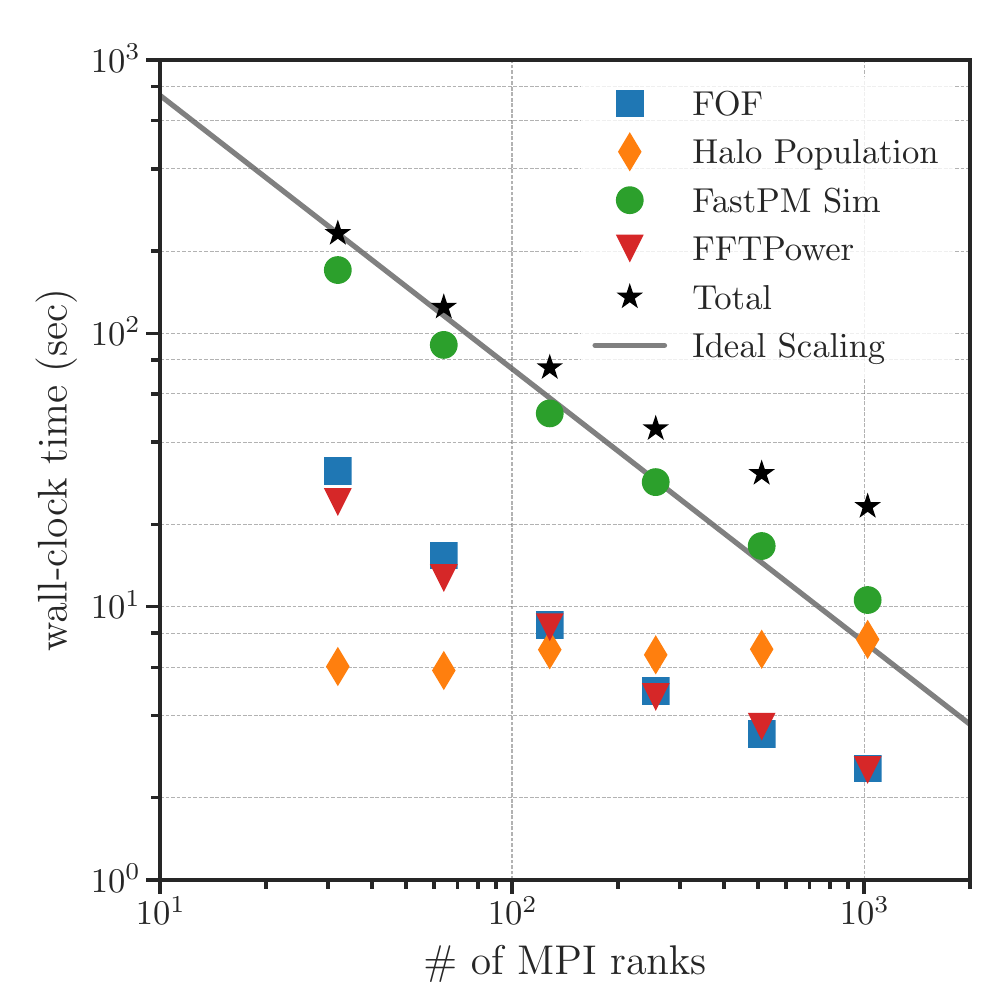}
\caption{The wall-clock time as a function of the number of MPI ranks used for each step in
the galaxy clustering emulator detailed in Figure~\ref{fig:emulator-code}.
Overall, the application shows excellent scaling behavior, with deviations
from the ideal scaling caused by the halo population step. This step does not
currently have a massively parallel implementation and takes a
roughly constant amount of time as more cores are used.
The benchmarks were performed on the NERSC Cori Phase I Haswell
nodes using 32 MPI ranks per node.}
\label{fig:emu-benchmark}
\end{figure}

In this section, we present performance benchmarks for several \nbkit{}
algorithms, as well as the emulator application discussed in
Section~\ref{sec:inaction}. Tests are run on the NERSC Cori Phase I
Haswell nodes, with 32 MPI cores per node. In Figure~\ref{sec:benchmarks}, we
show the strong scaling results for the \code{FFTPower}, \code{ConvolvedFFTPower},
\code{SimulationBoxPairCount}, and \code{SimulationBox3PCF} algorithms.
The benchmarks are performed for two different data configurations, meant
to simulate the data sets of current and future surveys, denoted
as ``small'' and ``large'', respectively. The ``small'' sample
is modeled after the completed BOSS galaxy sample
\citep{Reid:2016} and includes $10^6$ galaxies in a cubic box of side
length $L = 2500 \ h^{-1}\mathrm{Mpc}$. The ``large'' sample
includes a factor of 10 more objects in a box
of side length $L = 5000 \ h^{-1}\mathrm{Mpc}$ and is meant
to represent data from future surveys such as DESI \citep{DESI:2016}.
We run four sets of benchmarking tests:

\begin{itemize}
  \item \code{FFTPower}: compute $P(k,\mu)$ for 10 $\mu$ bins, using a mesh
  size of $N_\mathrm{mesh} = 1024$. This requires a single FFT operation.
  \item \code{ConvolvedFFTPower}: compute multipoles $P_\ell(k)$ for
  $\ell=0$, $2$, and $4$ for survey data (RA, Dec, $z$), using a
  mesh size of $N_\mathrm{mesh} = 1024$. The algorithm requires $2\ell+1$
  FFT operations per multipole, and 15 in total for this test.
  \item \code{SimulationBoxPairCount}: count the number of pairs as
  a function of separation for
  10 separation bins ranging from $r=10 \hMpc$ to $r=150 \hMpc$ and 100
  $\mu$ bins.
  \item \code{SimulationBox3PCF}: compute the isotropic 3PCF multipoles
  for $\ell=0,1,...,10$ and 10 separation bins ranging from $r=10 \hMpc$ to
  $r=150 \hMpc$.
\end{itemize}

In general, these four algorithms show excellent strong scaling with
the number of MPI ranks. For the power spectrum
algorithms (top row of Figure~\ref{fig:benchmarks}), the
dominant calculation is the FFT operation,
which has good scaling behavior.
Because the FFT is the dominant time cost, we find
nearly identical performances for the ``small'' and ``large'' samples.
The wall-clock time for the \code{ConvolvedFFTPower} algorithm is roughly
fifteen times that of the \code{FFTPower} algorithm, which is driven by
the total number of FFTs that each algorithm computes.
The pair-counting-based algorithms both take $\mathcal{O}(N^2)$ time to
compute their results. However, the \code{SimulationBoxPairCount} algorithm
relies on the highly optimized \Corrfunc{} software, which is significantly
faster than \code{SimulationBox3PCF}, which relies on $\kdcount{}$.
When using \code{SimulationBoxPairCount} on the ``small'' data set,
we find that MPI communication costs are non-negligible
due to the relatively small sample size, which hinders the scaling
performance of the code.

We also present performance benchmarks for the emulator application
described in Section~\ref{sec:inaction}.
For this test, we run a FastPM particle mesh simulation with $512^3$ total
particles. The halo finder
identifies roughly 225,000 dark matter halos that are then used to build a
mock galaxy catalog. The wall-clock times for
each step in the emulator are shown in Figure~\ref{fig:emu-benchmark}.
We see that the dominant fraction of the
wall-clock time is spent in the FastPM step, which
shows excellent strong scaling behavior up to the number of cores we
have tested. The implementation of the halo population step using
\halotools{} is not massively parallel, and therefore,
the time to solution for this step remains relatively constant as the
number of cores increases. The wall-clock time for this step only
becomes significant as the number of cores approaches $\sim$1024, and
it would be worth investigating improving this step's scaling if
users wish to run often at this scale. However, in our experience, we have
not found that the time cost of this step justifies
further efforts converting it to a massively parallel implementation.

We emphasize that for all benchmarks presented in this section, the number of
MPI ranks can always be increased such that the time to solution is on the
order of seconds. This becomes important for realistic data analyses in LSS,
which often require repeating an algorithm's calculation hundreds to thousands
of times, e.g., while sampling a parameter space using
Markov Chain Monte Carlo or optimization techniques. Due to the availability of large-scale
computing resources and the scaling behavior of \nbkit{} demonstrated here,
we believe that \nbkit{} will be able to meet the computational demands of
future LSS data analyses.

\section{Conclusions}
\label{sec:conclusions}

We have presented the first public release of \nbkit{} (v0.3.0), a massively
parallel Python toolkit for the analysis of large-scale structure
data. Relying on the \mpipy{} binding of MPI, the package
includes parallel implementations of a set of canonical algorithms
in the field of large-scale structure, including two and three-point
clustering estimators, halo identification and population tools, and
quasi-$N$-body simulation schemes. The toolkit also includes a set of
distributed data containers that support a variety of data formats
common in astronomy, including CSV, FITS, HDF5, binary, and \bigfile{} data.
With these tools, we hope \nbkit{} can serve as a foundation for the community
to build upon as we strive to meet the demands of future LSS data sets.

In designing \nbkit{}, we have attempted to balance the requirements
of both a scalable and interactive piece of software.
Our ultimate goal was to produce a piece of software that is as usable
in a Jupyter notebook environment as on an HPC machine.
We have adopted a modular, component-based approach that should enable
researchers to integrate \nbkit{} with their own software to build
complicated applications. As an illustration of its power, we have discussed an
implementation of a galaxy clustering emulator using \nbkit{},
which provides a complete forward model for the galaxy power spectrum,
starting from an initial, Gaussian density field.
We have also demonstrated that the toolkit shows excellent scaling behavior,
presenting a set of performance benchmarks for the
emulator as well as some of the more commonly used algorithms in
\nbkit{}.

We have outlined our development workflow for producing
a piece of reusable scientific software. \nbkit{} is open-source---we
strongly believe in the idea of open science and have placed an emphasis on
reproducibility when designing \nbkit{}. Designed for the LSS community,
we hope that new users will find \nbkit{}  useful in their own research
and that the software can continue to grow and mature in new ways from
community feedback and contributions. We are also strong believers in
the necessity of unit testing and adequate documentation for open-source
tools. We have attempted to meet these goals using the Travis automated
testing service and the Read the Docs documentation hosting tools.
Finally, we have aimed to
make \nbkit{} widely available and easily installable. The package supports
both Python versions 2 and 3, and binary distributions of \nbkit{} and its
dependencies can be installed onto Mac OS X and Linux machines using the
Anaconda package manager.

In the future, we hope to incorporate the expertise of new developers,
from both the LSS and Python HPC communities. We expect
the knowledge of both communities to be necessary in the data analysis of
future surveys. The set of features currently implemented in \nbkit{}
is not meant to be all-inclusive but rather a sampling of the more commonly
used tools in the field. Most importantly, we hope that \nbkit{} provides
a solid basis for the community to build upon. We warmly welcome feedback and
contributions of all forms from the community.
As open-source software, \nbkit{}
was designed as a tool to help the LSS community, and we hope that community
contributions can help maximize its benefits for its users.

\acknowledgments

NH and YF thank Martin White for comments on the design of the
correlation function algorithms and Manodeep Sinha and
Andrew Hearin for coordinating the software interfaces of \Corrfunc{} and
\halotools{} with \nbkit{}. NH and YF thank Rollin Thomas and Lisandro
Dalcin for discussions on MPI and Python on massively parallel HPC systems.
NH and YF thank Matthew Rocklin and Steven Hoyer for discussions on
building applications with \dask{}.
YF thanks Matthew Turk for insightful discussions about the design of
\textsf{yt}. We would also like to thank the communities supporting
the open-source software and tools that were invaluable to this
work: \numpy{}, \scipy{}, \pandas{}, \ipython{}, \jupyter{},
GitHub, Read the Docs, Travis, and Coveralls.
We are grateful for the suite of tools provided by Anaconda, a
trademarked Python binary distribution system for
scientific computing. We also thank Ray Donnelly and Mike Sarahan of
Continuum Analytics, Inc. for their help on building \nbkit{} binary packages.

In addition, a large number of researchers in the field of cosmology
provided useful feedback and input on the development of \nbkit{}:
Man-yat Chu,
Biwei Dai,
Zhejie Ding,
Lukas Heizmann,
Zvonimir Vlah,
Elena Massara,
Mehdi Rezaie,
Marcel Schmittful,
Hee-Jong Seo, and
Miguel Zumalac\'{a}rregui.

This work used resources of the National Energy Research Scientific Computing
Center, a DOE Office of Science User Facility supported by the Office of
Science of the U.S. Department of Energy under Contract No. DE-AC02-05CH11231.
NH is supported by the U.S. Department of Energy, Office of Science, Office of
Workforce Development for Teachers and Scientists, Office of Science Graduate
Student Research (SCGSR) program. The SCGSR program is administered by the Oak
Ridge Institute for Science and Education for the DOE under contract number
DE-SC0014664. Support for this work was also provided by the National
Aeronautics and Space Administration through Einstein Postdoctoral Fellowship
Award Number PF7-180167 issued by the Chandra X-ray Observatory Center, which
is operated by the Smithsonian Astrophysical Observatory for and on behalf of
the National Aeronautics Space Administration under contract NAS8-03060.
ZS also acknowledges support from a Chamberlain Fellowship at Lawrence Berkeley
National Laboratory (held previously to the Einstein) and from the Berkeley
Center for Cosmological Physics.
FB acknowledges support by an STFC Ernest Rutherford Fellowship,
grant reference ST/P004210/1.

\bibliographystyle{aasjournal}
\bibliography{main}

\end{document}